\documentclass[twocolumn,showpacs,preprintnumbers,amsmath,amssymb]{revtex4}

\usepackage{graphicx}

\usepackage{bm}

\begin{document}

\preprint{}

\title{Test of OPE and OGE through mixing angles of negative parity \\ $N^*$
resonances in electromagnetic transitions}

\author{He Jun}
\email{hejun@ihep.ac.cn}
\author{Dong Yu-bing}%

\affiliation{%
Institute of High Energy Physics, Chinese
Academy of Sciences, Beijing,
P.R.China
}%

\date{\today}

\begin{abstract}
In this report, by using the mixing angles of one-gluon-exchange
model(OGE) and one-pion-exchange model(OPE), and by using the
electromagnetic Hamiltonian of Close and Li, we calculate the
amplitudes of $L=1$ $N^*$ resonances for photoproduction and
electroproduction. The results are compared to experimental data.
It's found that the data support OGE, not OPE.
\end{abstract}

\maketitle

Which is the interaction between quarks mediated by, glouns or
mesons? In one form or another, it has been used in a wide variety
of models for the last two decades. In 2000, Isgur published his
critique\cite{critique} to the review\cite{GR} of Glozman and
Riska in which it is proposed that baryon spectroscopy can be
described by OPE without the standard OGE forces of ref. \cite{IK}
and \cite{OGE}.

In the critique, it is said that predicting the spectrum of baryon
resonances is not a very stringent test of a model. A prototypical
example is properties of the two $N^*_{1/2^-}$ states. Among
models which perfectly describe the spectrum, there is still a
composition of these states since all values of $\theta_{{1/2}^-}$
from 0 to $\pi$ correspond to distinct states. OPE model predicts
$\theta_{1/2^-}=\pm13^\circ$ and $\theta_{3/2^-}=\pm8^\circ$. Such
a $\theta_{1/2^-}$ has almost no impact on explaining the
anomalously large N$\eta$ branching ratio of the
$N^*(1535)_{1/2^-}$ and the anomalously small N$\eta$ branching
ratio of the $N^*(1650)_{1/2^-}$.

Recently, by using the method of Isgur and Karl \cite{IK77},
Chizma and Karl gave another values of mixing angles of OPE,
$\theta_{1/2^-}=25.5^\circ$ and $\theta_{3/2^-}=-52.7^\circ$
\cite{CK}. Their results are independent of spectrum and decay
data. We know that most values of mixing angles including the
"experiment" one are obtained through fitting the spectrum and
decay. The error of "experiment" values of Hey $et.al.$
\cite{expangle} is of order of $10^\circ$. Thus, we can't judge
which model is better through only comparing with "experiment"
mixing angle values. Since the predicting of the spectrum of
states is not enough to test a model \cite{critique}, it is
necessary to examine the exchange models with further experimental
data. Here, we will compare OPE and OGE in eletroproduction and
photoproduction through the mixing angles. Using the wave
functions obtained by Chizma and Karl, we calculate the amplitudes
of transition from ground state to $L=1$ $N^*$ resonances, then
compare results with experiment to test different models.

In ref. \cite{CK}, Chizma and Karl used the OGE and OPE
interaction Hamiltonians as following:
\begin{eqnarray}H_{hyp}^{OGE}=&&A\{(8\pi/3)\
 {\bf S}_1
\cdot{\bf S}_2\ \ \delta^3({\bf \rho})\ \ \ \ \nonumber\\&&+ (3{\bf
S}_1\cdot\widehat{{\bf \rho}}\ {\bf S}_2\cdot\widehat{{\bf
\rho}}-{\bf S}_1 \cdot{\bf S}_2)\  \rho^{-3}\}\ ,\\
 H_{hyp}^{OPE}=&&B\{(-4\pi/3)\
 {\bf S}_1
\cdot{\bf S}_2\ \ \delta^3({\bf \rho})\nonumber\\&&+ (3{\bf
S}_1\cdot\widehat{{\bf \rho}}\ {\bf S}_2\cdot\widehat{{\bf
\rho}}-{\bf S}_1 \cdot{\bf S}_2)\ \rho^{-3}\}\ {\bf
\lambda}_1^f\cdot{\bf \lambda}_2^f\ ,\end{eqnarray}where,
  $\hat{\rho}=\frac{\overrightarrow{\rho}}{|\overrightarrow{\rho}|}$. A, or
  B,
  is an overall constant which determines the strength of the interaction.
  ${\bf S}_{1,2}$, $\lambda^f_{1,2}$ are spins and the eight $3\times3$ Gell-Mann SU(3) flavor
  matrices for quarks number 1 and 2. Here we assumed the mass of pion is zero because
it does not change the results significantly.

Ignoring the color wavefunction, the harmonic-oscillator
wavefunctions for $L=1$ $N^*$ resonances have following
forms:\cite{IK}
\begin{eqnarray}S=3/2:&\ &
\Psi(^4 P)=\frac{1}{\sqrt{2}}\chi^s\{\psi^\lambda \phi^\lambda
+\psi^\rho \phi^\rho\}\ , \\ S=1/2:&\ & \Psi(^2
P)=\frac{1}{2}\{\chi^\lambda \psi^\rho \phi^\rho +\chi^\rho
\psi^\lambda \phi^\rho+\chi^\rho \psi^\rho \phi^\lambda
\nonumber\\&&-\chi^\lambda \psi^\lambda \phi^\lambda\}\ .\end{eqnarray}

The spin angular momentum $S=1/2$ or $3/2$ has to be coupled with
the orbital angular momentum $L=1$ to give the total angular
momentum $|L+S|\geq J\geq|L-S|$. As a result there are two states
each at $J=1/2$ and $J=3/2$, namely spin doublet and spin quartet:
$^2P_{1/2}$,$^4P_{1/2}$ and $^2P_{3/2}$,$^4P_{3/2}$. The physical
eigenstates are linear combinations of these two states, and can
be obtained by diagonalizing the Hamiltonian in this space of
states.
 Then, mixing angles are :
\cite{CK} \begin{eqnarray}&&OPE:\ \theta_{3/2}=-52.7^\circ, \
\theta_{1/2}=25.5^\circ \ ;\nonumber\\&&OGE:\
\theta_{3/2}=6^\circ,\ \theta_{1/2}=-32^\circ\ ,\end{eqnarray}and
the wave functions have the forms:
\begin{eqnarray}&&|N1700\rangle=cos\theta_{3/2}|^4P_{3/2}\rangle+
sin\theta_{3/2}|^2P_{3/2}\rangle\ ,\nonumber\\&&
|N1520\rangle=-sin\theta_{3/2}|^4P_{3/2}\rangle+
cos\theta_{3/2}|^2P_{3/2}\rangle\ ,
\\&& |N1650\rangle=cos\theta_{1/2}|^4P_{1/2}\rangle+
sin\theta_{1/2}|^2P_{1/2}\rangle\ ,\nonumber\\&&
|N1535\rangle=-sin\theta_{1/2}|^4P_{1/2}\rangle+
cos\theta_{1/2}|^2P_{1/2}\rangle\ .\end{eqnarray}

To calculate the electromagnetic transition amplitudes, we use the
electromagnetic interaction of Close and Li \cite{CL} which can be
derived from B-S equation \cite{brodsky}. It avoids the explicit
appearance of the binding potential through the method of McClary
and Byers \cite{MB}. Its explicit form
is:\begin{widetext}\begin{eqnarray} H^{em}&=&\sum_{i=1}^3
H_i=\sum_{i=1}^3 \{-e_i {\bf r}_i \cdot{\bf E}_
i+i\frac{e_i}{2m^*} ({\bf p}_i\cdot{\bf k}_i{\bf r}_i\cdot {\bf
A}_i+{\bf r}_i\cdot{\bf A}_i {\bf p}_i\cdot{\bf k}_i)- \mu_i{\bf
\sigma}_i{\bf B}_i\nonumber\\&&
-\frac{1}{2m^*}(2\mu_i-\frac{e_i}{2m^*})\frac{{\bf \sigma}_i}{2}
\cdot[{\bf E}_i\times{\bf p}_i-{\bf p}_i\times {\bf E}_i]\}+
\sum_{i<j}\frac{1}{2M_Tm^*}(\frac{{\bf \sigma}_i}{2} -\frac{{\bf
\sigma}_j}{2})\cdot[e_j{\bf E}_j\times{\bf p}_i- e_i{\bf
E}_i\times{\bf p}_i]\ ,
\end{eqnarray}\end{widetext}
where we keep to $ \emph{O}\rm$ $(1/m^2)$, and use long wave
approximation. ${\bf E}_i$ and ${\bf B}_i$ are the electromagnetic
fields, $e_i$, ${\bf \sigma}_i$, $\mu_i$ are the charge, spin, and
magnetic moment of quark $i$. $M_T$ is recoil mass. $m^*_i$ is the
effective quark mass including the effect of long-range scalar
simple harmonic potential, but it is independent on the exchange
potential. So $\mu_i$ or $m_i^*$ in two models can be
treated as the same free parameters.

By insertion of the usual radiation field for the absorption
of a photon into Eq.(8), and by integrating over the baryon
center-of-mass coordinate, we obtain the transverse photoexcited
value over flavor spin and spatial coordinates \cite{C}
\begin{equation}A_\lambda^N =\sum_{i=1}^3 \langle
X;J\lambda|H_i|N;\frac{1}{2}\lambda-1\rangle\ .\end{equation} Here
the initial photon has a momentum $\mathbf{k}||\widehat{z}$.  A
simple procedure, that of transforming the wave functions to a
basis which has redefined Jacobi coordinates, allows the
calculation of the matrix elements of the $H_1$ and $H_2$
operators to proceed in an exactly similar way to that of the
operator $H_3$. Calculation of the matrix elements of $H_3$ avoids
complicated functions of the relative coordinates in the
''recoil'' exponential.

By using the wave functions (6), (7), or non-admixture wave
functions (which can be seen as the wave functions with zero
mixing angles ) and by using Hamiltonian (8), we calculate the
amplitudes of photoexitation from the ground states N(p,n) to the
resonance X by Eq. (9) in Breit-frame. In the calculation, we
follow the convention of Koniuk and Isgur \cite{KI}. For the
photocouplings of the states made of light quarks, and the states
which are not highly exited, it should be a reasonable
approximation to treat the quark kinetic mass $m^*_i$ as a
constant effective mass $m^*$. As the reference \cite{C}, We keep
recoil mass at $M_T=3m^*$, and use parameter values, $\alpha=0.5
GeV$, and $m^*=0.437 GeV$. (in fact, the result isn't sensitive to
the values of $M_T$ and $m^*$)

In the photoproduction, nucleon is excited by a real photon, which
mass equals to zero, $i.e.$ $Q^2=0$ (here $Q^2=-q^2$,where $q^\mu$
is the transferred four-momentum). A useful measure of the quality
of the fit is to form a $\chi^2$ statistic in the usual way.
Introducing a "theoretical error" \cite{error} avoids overemphasis
in the fitting procedure of a few very well-measured
photocouplings. In Table I, we give amplitudes and $\chi^2$ of
non-admixture, of OPE, and of OGE, and list the experimental
values in last column.

\begin{table*}
\caption{ Breit-frame photoproduction amplitudes using wave
function of no-admixture(NA), of OPE, and of OGE. Here
$\alpha=0.5GeV$, $m^*=0.437GeV$, g=1.3, $M_T=3m^*$. Amplitudes are
in units of $10^{-3}Gev^{1/2}$; a factor of +i is suppressed for
all amplitudes. Experimental values are from PDG \cite{PDG}}
\renewcommand\tabcolsep{0.4cm}\begin{tabular}{ c  c  c c c c  c c r}  \hline\hline
 state & $A_\lambda^N$ & NA & $\chi^2_{NA}$& OPE & $\chi^2_{OPE}$& OGE & $\chi^2_{OGE}$&  Expt. \\\hline
  $N\frac{3}{2}^- (1700)$& $A^p_{\frac{1}{2}}$ & -21 & 0.0 & 29  & 3.9 & -26 & 0.1 & $ -18\pm13 $    \\
                         & $A^n_{\frac{1}{2}}$ &  19 & 0.1 & 33  & 0.4 &  17 & 0.1 & $   0\pm50 $    \\
                         & $A^p_{\frac{3}{2}}$ & -36 & 1.3 &-131 &17.2 & -21 & 0.4 & $  -1\pm24 $    \\
                         & $A^n_{\frac{3}{2}}$ & -14 & 0.1 &  89 & 3.6 & -27 & 0.2 & $  -3\pm44 $    \\
  $N\frac{3}{2}^-(1520)$ & $A^p_{\frac{1}{2}}$ & -23 & 0.0 & -31 & 0.1 & -21 & 0.0 & $ -24\pm9\ $    \\
                         & $A^n_{\frac{1}{2}}$ & -38 & 1.0 & -5  & 6.2 & -40 & 0.8 & $ -59\pm9\ $    \\
                         & $A^p_{\frac{3}{2}}$ & 139 & 1.8 & 55  & 29.0& 142 & 1.4 & $ 166\pm5\ $    \\
                         & $A^n_{\frac{3}{2}}$ &-125 & 0.4 & -74 & 8.1 &-124 & 0.4 & $-139\pm11 $    \\
  $N\frac{1}{2}^-(1650)$ & $A^p_{\frac{1}{2}}$ &  19 & 1.8 & -35 &11.9 &  81 & 1.2 & $  53\pm16 $    \\
                         & $A^n_{\frac{1}{2}}$ & -1  & 0.2 & 36  & 3.1 & -46 & 1.1 & $ -15\pm21 $    \\
  $N\frac{1}{2}^-(1535)$ & $A^p_{\frac{1}{2}}$ & 109 & 0.3 & 106 & 0.2 &  82 & 0.0 & $  90\pm30 $    \\
                         & $A^n_{\frac{1}{2}}$ & -82 & 1.1 & -75 & 0.8 & -66 & 0.4 & $ -46\pm27 $    \\
\hline \hline\end{tabular}
\end{table*}

In the first two columns of Table I, the amplitudes without
admixture and $\chi^2$ of those amplitudes are displayed. We can
see that amplitudes of many states agree with experimental data
well. But $A_{1/2}^p$ for $N(1650)$, $A_{3/2}^p$ for $N(1520)$,
$A_{1/2}^n$ for $N(1535)$ and $A_{3/2}^p$ for $N(1700)$ should be
uplifted. $A_{3/2}^n$ for $N(1520)$ and $A_{1/2}^n$ for $N(1520)$
should be suppressed. Obviously, if we mix two spin-1/2 states and
spin-3/2 states separately as Equs. [6], and [7], we can realize
it. The other noteworthy information we can get from the first two
columns is that the difference between $A_{1/2}^{p,n}$ for
$N(1650)$ and $A_{1/2}^{p,n}$ for $N(1535)$, or the difference
between $A_{3/2}^{p,n}$ for $N(1700)$ and $A_{3/2}^{p,n}$ for
$N(1520)$, is too large. So the admixture should not be very
large. Otherwise the results which have agreed with experiment
will be destroyed. The third and forth columns in Table I give the
results of OPE. Except $A_{1/2}^{p,n}$ for $N(1535)$, $\chi^2$ of
most amplitudes increase. $\chi^2$ of $A_{1/2}^p$ for $N(1650)$,
$A_{3/2}^{p}$ for $N(1700)$, or $A_{3/2}^{p}$ for $N(1520)$, is
even larger than 10. All those amplitudes are obtained by mixing
two amplitudes with large difference. Even some states change to
wrong direction. For example, $A_{1/2}^p$ for $N(1650)$ should be
uplifted, but admixture of OPE makes it lower. The sum $\chi^2$ of
twelve amplitudes also increases from 8.0 to 84.6. The fifth and
sixth columns present results of OGE. Admixture of OGE gives
significant improvement on no-admixture results. Almost all
amplitudes agree with experiment well. The sum value of $\chi^2$
decreases from 8.0 to 6.2.

Electroproduction amplitudes are extracted from eN scattering. In
this procedure, nucleon is excited from ground state to excited
state by a virtual photon, which mass isn't zero, $i.\ e.$
$-Q^2\neq 0$. In Fig 1, We draw curves of calculated amplitudes,
which vary with $Q^2$. The results of no-admixture, OPE, and OGE
are presented along with experimental data in the figure.

In Fig. 1 (b), $A_{1/2}^{p}$ for $N(1535)$, and Fig. 1 (d),
$A_{1/2}^{p}$ for $N(1520)$, the differences between OPE and OGE
are small. The relativistic effect on wave functions, which we did
not consider in this paper, may smear the small differences. So
they are useless to compare OPE and OGE. Discrepancies of
different models in the other two graphs are large. In Fig 1 (a),
$A_{1/2}^{p}$ for $N(1650)$, and (c), $A_{3/2}^{p}$ for $N(1520)$,
OGE is superior to OPE obviously. In addition, we can see that in
Fig. 1 (a) the curve without admixture is between those of OPE and
OPE. It suggest that one of models will give wrong direction
correction. According to data and our results of photoproduction,
it should be OPE. In Fig. 1 (c) OPE gives too large correction
obviously.

Though we use the non-relativistic wave functions here, from Table
I of reference \cite{C} and from the calculations of this paper,
we can find that relativistic effect won't reverse our conclusion.
For the most results with large differences between OPE and OGE,
the conclusion can be kept when we change mixing angles of OPE and
OGE separately by $\pm10^\circ$. For example, the sum of $\chi^2$
for OGE varies between 5.1 and 11.3, and that of OPE varies
between 57.5 and 117.5. In this case, OGE is still superior to OPE
obviously. Through our calculation, it is believable that the OGE
is better than OPE in fit with photoproduction and
electroproduction amplitudes of the $N^*$ resonances with negative
parity. In other words, OGE gives consistent mixing angles to
explain spectrum, decay branching, photoproduction and
electroproduction amplitudes.

\

\begin{acknowledgments}
This work is supported by the National Natural Science Foundation
od China No. 10075056, by CAS Knowledge Innovation Project No.
KL2-SW-N02.
\end{acknowledgments}

\begin{figure*}
  \includegraphics[bb=65 385 530 770,clip=true]{graph.prn}\\

\caption{Breit-frame electroproduction amplitudes. Here
$\alpha=0.5GeV$, $m^*=0.437GeV$, g=1.3, $M_T=3m^*$. A factor of +i
is suppressed. Full curves are calculated in OGE, dashed ones in
OPE, and dotted ones without admixture. Experimental values are
from \cite{WSWR, data, PDG}}

 \label{fig}
\end{figure*}

\end{document}